\documentclass[12pt,preprint]{aastex}

\shorttitle{Embedding Lagrangian Sink Particles in Eulerian Grids}
\shortauthors{Krumholz, Klein, \& McKee}

\newcommand{\calm}{\mathcal{M}}

\newcommand{\rb}{r_{\rm B}}
\newcommand{\tb}{t_{\rm B}}
\newcommand{\rbh}{r_{\rm BH}}
\newcommand{\tbh}{t_{\rm BH}}

\begin{document}

\title{Embedding Lagrangian Sink Particles in Eulerian Grids}

\author{Mark R. Krumholz}
\affil{Physics Department, University of California, Berkeley,
Berkeley, CA 94720}
\email{krumholz@astron.berkeley.edu}

\author{Christopher F. McKee}
\affil{Departments of Physics and Astronomy, University of California,
Berkeley, Berkeley, CA 94720}
\email{cmckee@astron.berkeley.edu}

\author{Richard I. Klein}
\affil{Astronomy Department, University of California, Berkeley,
Berkeley, CA 94720, and Lawrence Livermore National Laboratory,
P.O. Box 808, L-23, Livermore, CA 94550}
\email{klein@astron.berkeley.edu}

\begin{abstract}
We introduce a new computational method for embedding Lagrangian sink
particles into an Eulerian calculation. Simulations of gravitational
collapse or accretion generally produce regions whose density
greatly exceeds the mean density in the simulation. These dense
regions require extremely small time steps to maintain numerical
stability. Smoothed particle hydrodynamics (SPH) codes approach this
problem by introducing non-gaseous, accreting sink particles, and
Eulerian codes may introduce fixed sink cells. However, until now
there has been no approach that allows Eulerian codes to follow
accretion onto multiple, moving objects. We have removed that
limitation by extending the sink particle capability to Eulerian
hydrodynamics codes. We have tested this new method and found that it
produces excellent agreement with analytic solutions. In analyzing our 
sink particle method, we present a method for evaluating the disk 
viscosity parameter $\alpha$ due to the numerical viscosity of a hydrodynamics
code, and use it to compute $\alpha$ for our Cartesian AMR code. We also present
a simple application of this new method: studying the transition
from Bondi to Bondi-Hoyle accretion that occurs when a shock hits a
particle undergoing Bondi accretion.

\end{abstract}

\keywords{Accretion, accretion disks --- hydrodynamics --- methods:
numerical --- shock waves}

\section{Introduction}

Simulations of gaseous collapse and accretion are ubiquitous in
astrophysics, and all of them face a common difficulty: gravitational
collapse leads to the formation of structures on length scales very
small compared to the initial collapsing object. This leads to an
enormous dynamic range that generally makes the full problem
computationally infeasible. Simulations of star formation, for
example, generally start with observed molecular cloud cores that are
$\sim 0.1-1.0$ pc in size \citep{williams2000}, while the stars that
are the endpoints of the calculation are of order a solar radius
($\sim 10^{11}$ cm) in size -- a dynamic range of $\sim 10^7$ in
length.

Dynamic range is expensive for two distinct reasons. First, one
requires enough resolution elements (cells for gridded codes,
particles for gridless codes) to resolve both the largest and smallest
structures in the problem. For an Eulerian code with no adaptivity,
such as the widely used ZEUS package \citep{zeus1, zeus2, zeus3},
increasing the linear resolution of a calculation by a factor $f$
requires increasing the number of cells in each dimension $f$, thereby 
increasing the total number of cells by a factor of $f^N$, where $N$
is the number of dimensions. Lagrangian approaches, such as SPH
\citep{sph1,sph2} and adaptive Eulerian approaches, such as AMR
\citep{berger84, berger89, bell94}, fare significantly better in this
regard by following the mass and adding resolution elements only in
regions of interest.

The second reason a large dynamic range is expensive is that the
smallest and largest structures present in a problem often evolve on
very disparate time scales, requiring a simulation to take an
inordinate number of very small time steps. In explicit hydrodynamics codes
this problem is embodied in the Courant condition \citep{richtmyer67},
which requires that the time step be less than the signal-crossing
time of a resolution element. Increasing the linear resolution by a
factor $f$ therefore generally requires multiplying the number of time
steps by $f$ as well. Again, adaptive methods that allow different
time steps for different resolution elements \citep{bate95, fisher04}
do somewhat better.

Even with the improvements in computational efficiency made possible
by adaptivity, however, many interesting problems require more dynamic
range than any code can handle. The time stepping constraint in
particular is a significant barrier because, unlike additional cells,
the additional time steps cannot easily be distributed on a parallel
machine. Therefore simulators have introduced sinks: regions of a
flow that accrete incoming material but that have no internal
structure and therefore no requirements for high resolution in either
time or space. Sinks provide a way to stop following collapse at a
pre-chosen scale (hopefully) without damaging the rest of the
calculation, thereby preventing the time step from grinding to zero
and the number of resolution elements from running off to infinity.

\citet{bate95} introduced the technique of sink particles in SPH
codes. Sink particles are perfect absorbers that accrete
other particles that approach within a certain distance. The accretion 
radius sets the smallest scale that the simulation can resolve. The
boundary pressure of a sink particle is determined by extrapolation
from the particles around them. While this technique has proven
extremely useful and has been adopted in numerous places, SPH codes
are not appropriate for all problems. To date there is no widely-used,
well-tested, SPH code that includes magnetohydrodynamics or radiative
transfer (although see \citet{price2003a, price2003b} for a recent
implementation of MHD with SPH that is still in the testing phase). In
addition, SPH codes require significant artificial
viscosity, which tends to cause artificial heating and smearing at
shocks and interfaces. SPH is therefore of limited utility in
problems where these features are important \citep{shapiro96}.

In contrast, Eulerian codes including MHD or radiative transfer are
reasonably mature, and, using Godunov schemes, require much less artificial
viscosity \citep{truelove98}. Eulerian codes have included sinks in the 
form of sink cells \citep{bb82}. Similar to sink particles, sink cells
allow mass to enter but not to leave, and their boundary pressures are
found by extrapolation from neighboring cells. The disadvantage to this
approach is that sink cells are fixed in the grid and are therefore
inapplicable in cases where there are multiple accretors moving
relative to one another (for example a binary system) or where one
does not know in advance where an accretion center will form (for
example in star formation in a turbulent medium).

In this paper we introduce a technique to embed a Lagrangian sink 
particle in an Eulerian code. This technique allows us to use Eulerian 
codes for cases where they are preferred, while retaining the
flexibility of a moving sink center. While previous work has combined
non-fluid particles with Eulerian hydrodynamics
\citep{kravtsov03}, our approach is unique in that it allows the 
non-fluid particles to accrete from the gas, and therefore truly act
as sinks. In \S~\ref{methodology} we
introduce the Eulerian AMR code and describe the method we use to
embed Lagrangian sink particles within it. In \S~\ref{tests} we
describe tests that we have done to evaluate the accuracy of this
method. In discussing our test of a sink particle in the context of a
disk, in \S~\ref{diskalpha}, we estimate the standard viscosity
parameter $\alpha$ \citep{shakura1973}.
We then consider a simple application of our technique in
\S~\ref{bondi2bh}: modeling the process of a flow changing from
Bondi accretion to Bondi-Hoyle accretion as a result of an external
shock. Finally, we discuss our conclusions in
\S~\ref{conclusion}.

\section{Computational Methodology}
\label{methodology}

\subsection{The Eulerian Code}

For the calculations presented in this paper we used our 3-D AMR
code. The code includes hydrodynamics, gravity,
and radiative transfer, but we shall refer only to the first two
components in this paper. The hydrodynamics module solves the Euler
equations for a compressible, multi-fluid system,
\begin{eqnarray}
\frac{\partial\rho^i}{\partial t} + \nabla\cdot\left(\rho^i
\mathbf{v}\right) & = & 0 \\
\frac{\partial}{\partial t} \left(\rho^i \mathbf{v}\right) + \nabla
\cdot \left(\rho^i \mathbf{vv}\right) & = & -\nabla\sum_i P^i -
\rho^i\nabla\phi \\
\frac{\partial}{\partial t}\left(\rho^i e^i\right) + \nabla \cdot
\left[\left(\rho^i e^i + P^i\right)\mathbf{v}\right] & = & \rho^i
\mathbf{v} \cdot \nabla\phi,
\end{eqnarray}
where $\rho^i$ is the density of fluid $i$, $\mathbf{v}$ is the vector
velocity (taken to be the same over all fluids), $P^i$ is the
thermal pressure, and $e^i$ is the total non-gravitational energy per
unit mass. We determine the potential $\phi$ by solving the Poisson
equation as described below. The code solves these equations using a
conservative high-order Godunov scheme with an optimized approximate
Riemann solver \citep{toro97}. The algorithm is second-order accurate
in both space and time for smooth flows, and it provides robust
treatment of shocks and discontinuities.

The gravitational module solves the Poisson equation
\begin{equation}
\nabla^2\phi = 4\pi G \sum_i \rho^i
\end{equation}
on an adaptive grid hierarchy (as described below) to find the
gravitational potential from a given density distribution. In each
time step we compute the potential and then use it as a source term in
the hydrodynamics equations as shown above \citep{truelove98}. The
gravity module uses a multigrid iteration scheme to solve the
linearized Poisson equation on each level of the adaptive
hierarchy.

Each physics module operates within the AMR framework \citep{berger84,
berger89, bell94}. We discretize the problem domain onto a base, coarse
level, denoted level 0. We dynamically create finer levels, numbered
$1,2,\ldots n$, nested within that coarse
level as needed. The nesting process is recursive, so each fine level
may contain even finer levels, providing no theoretical upper limit to
the maximum resolution. In practice, limits of computational resources
require that we select a maximum level of refinement allowed for a
given calculation. The process for a time step is similarly
recursive: one advances level $0$ through a single time step $\Delta
t_0$, then advances each subsequent level for the same amount of 
time. Each level has its own time step, and in general
$\Delta t_{l+1}< \Delta t_l$, so after advancing level 0 we must
advance level 1 through several steps of size $\Delta t_1$, until it
has advanced a total time $\Delta t_0$ as well. At that point we apply 
a synchronization procedure to guarantee conservation of mass,
momentum, and energy across the boundary between levels 0 and 1.
However, each time we advance level 1 through time $\Delta t_1$, we
must advance level 2 through several steps of size $\Delta t_2$, and
so forth to the finest level present.

The sink particle framework we present below is not dependent on any
of the details of the AMR framework, and may be applied equally well
to fixed-grid codes. Therefore, we will suppress discussion of AMR
details. However, we note here that we set our refinement criteria to
guarantee that a sink particle's accretion zone (see
section~\ref{accretezone}) is always refined to the highest allowable
level for a given calculation. When we refer to cell spacings and time 
steps in what follows, the cell spacings are always those of the
finest AMR level. In addition, for simplicity we shall also suppress
all discussion of multi-fluid issues.

\subsection{Creation of Sink Particles}

We may either introduce sink particles in the initial
conditions for a calculation, or we may create them when
necessary. Once we introduce a sink particle, we lose all knowledge of 
the flow in some region around it and assume that the gas within that
region will continue to collapse beyond the scale resolved by our
simulation. We therefore wish to introduce sinks only when the
when there is good physical reason to believe that continuing the
calculation without the sink will give inaccurate results, and that
the gas in the vicinity of the sink is likely to continue collapsing
past the scales resolved in our calculation.

Both of these conditions are met in cells that
violate the Jeans criterion \citep{truelove97, truelove98}:
\begin{equation}
\label{jeanslength}
\Delta x < J \lambda_J = J \sqrt{\frac{\pi c_s^2}{G \rho}}.
\end{equation}
Here, $J$ is a constant of order unity, $\lambda_J$ is the Jeans
length in a cell of length $\Delta x$, and $c_s$ and $\rho$ are the
sound speed and density in the cell. \citet{truelove97} found that
$J=0.25$ is sufficient to prevent artificial fragmentation in most of
the problems they considered. We can also write this as a condition
on the density for fixed $\Delta x$,
\begin{equation}
\rho_J < J^2 \frac{\pi c_s^2}{G \Delta x^2}.
\end{equation}
If the density in any cell does exceed $\rho_J$ with $J=0.25$, we
create in the center of that cell a sink particle with mass
\begin{equation}
m_{\rm sink} = \left[\rho-\rho_J\left(0.25\right)\right] \Delta x^3,
\end{equation}
so the density of the gas remaining in the cell is
$\rho_J\left(0.25\right)$. We also transfer a proportional amount of
momentum and energy from the gas to the sink particle.

\citet{truelove97} have shown that
continuing a calculation that has violated the Jeans criterion will lead
to artificial fragmentation.  By creating sink particles in cells that
violate the Jeans condition we prevent this from occurring. 
Also, since the density in the cell must have
been lower in the time step before the sink particle was appeared, it
follows from continuity that $\nabla \cdot \mathbf{v} < 0$ in that
cell. The fact that the cell violates the Jeans criterion indicates
that its self-gravity has begun to become important.
It is therefore likely that the gas in that cell will continue
collapsing indefinitely. Thus, creating sinks in cells that violate
the Jeans criterion meets the conditions that we create sinks only
when necessary and only when we are confident that indefinite collapse
is a valid approximation to the true behavior of the system. Note that
these criteria are roughly analogous to those used by SPH codes
\citep{bate95, bcl02} to create sink particles: the particle must be 
in a region of converging flow that is gravitationally bound.

There is one cautionary note: for a cell to violate the Jeans
criterion, its mass must be at least
\begin{equation}
\label{cellmass}
m_{\rm cell} > \rho \Delta x^3 = \pi^{3/2} J^3 \frac{c_s^3}{\left(G^3
\rho \right)^{1/2}}
\end{equation}
where in the last step we eliminated $\Delta x$ using equation
(\ref{jeanslength}).
In comparison, the maximum mass possible for a stable,
self-gravitating, isothermal object is the Bonnor-Ebert mass
\citep{ebert, bonnor}, $m_{\rm BE} = 1.18 \; c_s^3 /\left(G^3
\rho\right)^{1/2}$.
A Jeans-violating cell (for $J=0.25$) therefore has a minimum mass of
\begin{equation}
m_{\rm cell} > \frac{\pi^{3/2}}{1.18} J^3 m_{\rm BE} \approx 0.07 \;
m_{\rm BE}.
\end{equation}
While the addition of neighboring cells will likely raise the total
mass in the region above the Bonnor-Ebert mass, it is therefore still
possible to create a sink particle in a region that is
stable against collapse. However, a newly created sink particle
has a very low accretion rate (see
section~\ref{accretion}). If the region is truly stable, gas will not
continue to accrete onto the sink, the sink's mass will remain very
low. We thus have an ex post facto check on the validity of our sink
particle creation method. If such a situation occurs, there is no way
to do the calculation at the chosen level of resolution without
violating the Jeans condition or creating a suspiciously small sink
particle. The only choice is to redo the calculation at a higher
resolution.

\subsection{Merging Sink Particles}
\label{merge}

In a region of gravitational collapse, we often find that in a single
time step a block of contiguous cells increases in density so that
they all violate the Jeans condition and create sink particles. When
this happens we wish to merge these particles, since allowing them to
remain unmerged and possibly separate would risk a solution that
contains resolution-dependent artificial fragmentation. In addition,
when a sink particle is first created, gas usually continues to flow
into the sink particle's host cell and its neighbors. Since a
newly-created sink particle's mass and rate of gas accretion (see
section~\ref{accretion}) are very low, after each time step or two these
cells will once again violate the Jeans condition. As a result, the
code will create more sink particles, which ought to be merged
into the already existing one. (This process generally continues until 
the sink particle is massive enough that its gas accretion rate
prevents the density in the host cell from rising above the Jeans
density.)

To deal with this phenomenon, at the end of each time step
we group all the sink particles present in the calculation
using a friends of friends (FOF) algorithm \citep{davis85} with a
linking length equal to the radius of the accretion zone (see
section~\ref{accretezone}). We then merge all groups of particles that
the FOF algorithm finds. We replace the merged group by a single
particle at the center of mass of the group, and we add all particle
quantities conservatively. Later on in the calculation, if two
independently formed objects pass near one another for a short period, 
we may not wish to merge them. In this case we may temporarily reduce
the radius of the accretion zone (which for technical reasons
must be no larger than the merger radius) during the close passage and 
increase it again once the objects are sufficiently far apart. Since
sink particles moving under gravity will spend very little time during
a close passage, this will have a negligible effect on the overall
accretion rate or final mass. This adjustment to the merger radius may 
be handled either manually or by an automated algorithm.

\subsection{Accretion onto Sink Particles}

\subsubsection{The Accretion Rate}
\label{accretion}

Once it appears, a sink particle accretes gas from the surrounding
cells. The gas in which the sink particle is embedded continues to
evolve according to the Euler equations. Setting the accretion rate
is therefore critical in cases where the flow
onto the sink particle is subsonic, because the rate at which mass
flows from Eulerian cells to the pressureless particle
determines the amount of back-pressure opposing the accretion
flow. The accretion formalism thus serves a function analogous to the
extrapolation procedure used to find boundary pressure in SPH sink
particle or Eulerian sink cell formalisms.

We characterize the relative importance of pressure versus gravity via
the particle's Bondi-Hoyle radius \citep{bondi52},
\begin{equation}
\label{rbh}
r_{\rm BH} = \frac{G M}{v_{\infty}^2+c_{\infty}^2},
\end{equation}
where $M$ is the particle's mass and $v_{\infty}$ and $c_{\infty}$ are
the velocity and sound speed of the gas far from the sink
particle.

In the limit where the Bondi-Hoyle radius is much larger than a
cell spacing, the choice of accretion rate and hence the pressure is
irrelevant because the flow is supersonic near the sink
particle. Even if the accretion rate is set too small, gas will flow into the
sink particle's host cell until its density is high enough to violate
the Jeans condition. Once that happens, the code will create new sink
particles that will immediately merge with the existing particle, thus 
setting an effective accretion rate higher than that given by the
formula. In the opposite limit, $r_{\rm BH} \ll \Delta x$, the sink particle
is simply a point mass moving through a uniform gas. Its gravity is
relevant only on scales smaller than we resolve in the
simulation. This is just the classical Bondi-Hoyle-Lyttleton problem,
which has analytic solutions in the limits $v_{\infty} \ll
c_{\infty}$ and $v_{\infty} \gg c_{\infty}$ \citep{hoyle39,
hoyle40a, hoyle40b, hoyle40c, bondi52}. In between these two limits,
\citet{ruffert94a} and \citet{ruffert94b} give the approximate formula
\begin{equation}
\label{accrate}
\dot{M} = 4 \pi \rho_{\infty} G^2 M^2 \left[ \frac{\lambda^2
c_{\infty}^2 + v_{\infty}^2}{\left(c_{\infty}^2+v_{\infty}^2\right)^4}
\right]^{1/2}
= 4\pi \rho_{\infty} r_{\rm BH}^2 \left(\lambda^2
c_{\infty}^2+v_{\infty}^2\right)^{1/2}.
\end{equation}
Here, $\lambda$ is a constant of order unity that depends on the
equation of state of the gas. For an isothermal gas, $\lambda =
e^{3/2}/4\approx 1.120$, and we use that value throughout this work

For a real simulation
with complex, turbulent flows, there is no obvious way to choose
$v_{\infty}$. For symmetric accretion flows, the gas in the sink
particle's host cell is generally co-moving with the background at
large distances. Since we have no better alternative, we therefore
take $v_{\infty}$ to be the relative velocity of the sink particle
and the gas in its host cell. Similarly, we use sound speed in the
host cell for $c_{\infty}$. The choice of $\rho_{\infty}$ requires
more discussion. We let
ourselves be guided by the behavior we expect when simulating simple
Bondi accretion. In that case, the density profile, which we denote
by $\alpha(x)\equiv\rho(x)/\rho_{\infty}$, is the solution to a pair of
coupled non-linear ordinary differential equations
\citep{bondi52}. Here, $x\equiv r/r_{\rm BH}$ is the dimensionless
radius. The density in the central cell should be $\sim
\rho_{\infty} \alpha\left(\Delta x/r_{\rm BH}\right)$. We therefore
set
\begin{equation}
\label{rhobar}
\rho_{\infty} = \frac{\overline{\rho}}{\alpha \left(1.2 \; \Delta
x/r_{\rm BH}\right)},
\end{equation}
where $\overline{\rho}$ is the weighted mean density in the accretion
region (see section~\ref{accretezone}). In the limit $\Delta x \gg
r_{\rm BH}$, $\alpha \left(1.2 \; \Delta x/r_{\rm BH}\right)
\rightarrow 1$, so we recover the correct behavior for this case. We
inserted the factor of 1.2 because we found that it gave improved
results in the intermediate range $\Delta x \sim r_{\rm BH}$ (see
section~\ref{bondisim}).

The accretion rate is reduced in the presence of rotation
as described in section~\ref{accretezone}.

As a final note, \citet{ruffert94c} gives a somewhat more complex
formula than (\ref{accrate}) that provides a slightly better fit to
accretion rates found in their numerical simulations. However, even
that formula is off by as much as 60\% for isothermal flows with
intermediate Mach numbers \citep{ruffert96}, and thus we decided
against the additional complexity involved in implementing it.
Fortunately, as we discuss in section~\ref{bondihoyle}, errors in the
accretion formula tend to be self-correcting in an actual simulation,
and thus the details of the accretion formula are not critical.

\subsubsection{The Accretion Zone}
\label{accretezone}

We wish the accretion rate to change smoothly as the sink particle
moves across cell boundaries. Therefore we define an accretion zone
around each sink particle. We set the accretion rate based on
average properties in the accretion zone, and when the sink particle
accretes mass, it does so from all cells within the accretion zone. In
choosing the size of the accretion zone, there are two competing
factors. Since the solution within the accretion zone is artificially
affected by the accretion process, the larger the accretion zone, the
larger the region in which we give up on the accuracy of the
solution. However, in order to compute accurately the rate at which
mass enters the accretion zone, we must have adequate resolution on
its boundary. In addition, the size of the accretion zone will
determine our ability to resolve anisotropies in the accretion flow.
We define the accretion region as all
the cells within a radius $r_{\rm acc}$ of the sink particle's host
cell. Based on experiments with different sizes, we adopt a value
$r_{\rm acc}=4\Delta x$ throughout this work. However, our
implementation of the sink particle algorithm leaves the radius of the
accretion region  as a free parameter to be set at run time.

In cases where the particle's Bondi-Hoyle radius is smaller than the
accretion zone, it would be incorrect to set the accretion rate rate
based on a uniform average of all cells, however. Therefore we define
an accretion kernel with radius
\begin{equation}
r_K = \left\{ \begin{array}{r@{\quad:\quad}l}
\Delta x / 4 & r_{\rm BH} < \Delta x/4 \\
r_{\rm BH} & \Delta x / 4 \le r_{\rm BH} \le r_{\rm acc}/2 \\
r_{\rm acc} / 2 & r_{\rm BH} > r_{\rm acc} / 2
\end{array}.
\right.
\end{equation}
Within the accretion zone we assign each cell a weight
\begin{equation}
\label{weight}
w \propto \exp\left(-r^2/r_K^2\right),
\end{equation}
where $r$ is the distance from the cell
center to the sink particle. The minimum value of $\Delta x / 4$ for
$r_K$ ensures that the accretion changes smoothly as the particle
crosses cell boundaries even when $r_{\rm BH}$ is small. The maximum
value of $r_{\rm acc}/2$ ensures that cells at the edge of the accretion
region have little weight and thus there are no sudden changes in the
accretion rate as cells enter or leave the accretion region. Once we
have assigned weights to all cells in the accretion zone, we use a
weighted average to set $\overline{\rho}$ in equation
(\ref{rhobar}). We then compute the accretion rate for this time step
from equation (\ref{accrate}). Note that this procedure becomes
undefined if the accretion zones of multiple sink particles overlap;
for this reason we require that the sink particle merger radius
always be greater than or equal to the accretion zone radius.

Thus far our algorithm has not included the effects of angular
momentum, which may substantially reduce the accretion rate relative
to the spherically symmetric case. Furthermore, in a rotating flow,
low angular momentum gas along the polar axis accretes more easily
than high angular momentum gas in the equatorial plane. Thus, it would
be incorrect to use an accretion algorithm that accretes equally
quickly from all cells regardless of their place in the rotating
flow. We therefore choose a strategy that will both reduce the
accretion rate in the presence of rotation and allow accretion to
occur anisotropically from within the accretion zone.

To include rotation, we first divide the mass to be transferred to 
the sink particle (computed via equation \ref{accrate}) among the
cells in the accretion zone so that each cell contributes an amount of
mass proportional to its weight (computed via equation
\ref{weight}). We then divide each cell into $8^3$ identical point
particles arranged in a uniform grid
throughout the cell, each with $1/8^3$ the mass, momentum, and energy
of the cell. For each point particle we compute 
its distance of closest approach to the sink particle if it were to
travel on a purely ballistic trajectory while the sink particle
moved at constant velocity. For a point particle with specific angular
momentum $j_{\rm sp}$ and specific energy $e_{\rm sp}$ (kinetic plus
gravitational) in the sink particle's rest frame, this distance is
\begin{equation}
r_{\rm min} = -\frac{GM}{2e_{\rm sp}} \; \left[1-
\sqrt{1+\frac{2 j_{\rm sp} e_{\rm sp}}{\left(GM\right)^2}}\,\right].
\end{equation}
If the point particle is not bound to the sink particle, so that
$e_{\rm sp}>0$, we take $r_{\rm min}=\infty$.
We do not want the sink particle to accrete material that has too much 
angular momentum to reach its ``surface''. We therefore count up the
number $n$ of point particles for which $r_{\rm min}>\Delta x/4$ and
reduce the amount of mass to be accreted from that cell by a factor
$n/8^3$. (The factor of 4 in $r_{\rm min}$ is to ensure that the sink
particle has an effective size smaller than the size of a cell;
experimentation with different values from 0.1 to 0.5 produced no
noticeable  differences in behavior). For the host cell, since we
cannot compute a meaningful
specific angular momentum, we set $n$ equal to the maximum of the
values of $n$ in the cells bordering the host cell if $r_K \ge
\Delta x/4$, or $n=0$ (i.e. uninhibited accretion) if $r_K<
\Delta x/4$. Once this is done
we subtract the appropriate amount of mass from each cell and add it
to the sink particle. To ensure stability, we also set an absolute cap
that no more than 25\% of the mass may be removed from a cell in any
single time step.

Next we must compute the amount of linear momentum the sink particle
accretes. Since it represents an object far smaller than the grid size
in the calculation, it should accrete negligible angular momentum and
exert no torques on the gas. We therefore divide each cell's momentum,
taken in the sink particle's rest frame, into components parallel and
transverse to the radial vector connecting that cell to the sink
particle.  We reduce the cell's radial momentum by a factor equal to
the fraction of the cell's mass that we have accreted, while we leave
its transverse momentum unchanged. Thus, accretion preserves the
{\it radial velocity} and the {\it angular momentum} of the gas. To
ensure linear momentum conservation, we change the sink particle's
momentum by an amount equal and opposite to the total change in the
momentum of the gas cells. Finally, we compute the new total energy of
each gas cell by keeping the cell's specific thermal energy constant
while computing a new kinetic energy based on its new density and
momentum.

We find that this accretion procedure conserves mass, momentum, and
angular momentum to machine precision. However, for more discussion of 
angular momentum conservation see section~\ref{rotflow}.

\subsection{Motion of Sink Particles}

In each time step we update the position of each sink particle based
on its current momentum, and we modify its momentum through accretion
(as described in Section~\ref{accretion}) and through gravity. For
reasons of algorithmic speed, we handle the
position and momentum update of sink particles in two steps. First, we
change the momenta of sink particles due to their gravitational
interactions with the the gas by an amount $F_{\mbox{\scriptsize
gas-part}} \Delta t$, where $\Delta t$ is the time step and
$F_{\mbox{\scriptsize gas-part}}$ is the gas-particle
gravitational force. To ensure accuracy, we constrain the time step to
require that
\begin{equation}
\label{partcourant}
\max\left(v_{\mbox{\scriptsize part}}\right) \Delta t < C\Delta x,
\end{equation}
where $\max\left(v_{\mbox{\scriptsize part}}\right)$ is the largest
particle velocity in the calculation, $C$ is a constant of order
unity (for our runs generally 0.5), and $\Delta x$ is a cell
spacing. This restriction is usually less stringent than the ordinary
gas Courant condition. Particles have velocities comparable to the gas 
out of which they form, and the collapsing gas from which sink
particles form is usually not the gas that has the highest velocity
relative to the grid.

To compute the force on a
particle, we use a Plummer law \citep{aarseth63},
\begin{equation}
\mathbf{F}_{\mbox{\scriptsize gas-part}} = -G m \int
\frac{\rho}{r^2+\epsilon^2} \frac{\mathbf{r}}{r} dV
\end{equation}
where $m$ is the mass of the particle, $\rho$ is the gas density,
$\mathbf{r}$ is the vector from the
sink particle to a given cell or particle, and $\epsilon$ is the
softening length, which we leave as a parameter that may be set at
runtime. In general, the softening length should be smaller than the
size of the accretion region, to ensure that softening does not alter
the rate at which gas crosses its boundary. Choosing a smaller
softening length, however, increases the maximum velocity the gas will 
attain within the accretion region as it falls onto a sink particle,
which will in turn decrease the time step due to the Courant
condition. For our default choice of $r_{\rm
acc}=4\Delta x$, we therefore set a default value of $\epsilon =
2\Delta x$. Since the number of
particles is generally small, we compute gas-particle forces via a
direct sum. In computing the force between a particle and the gas, we
treat all cells except the particle's host cell and its neighbors as
point masses 
located at the cell center. To compute the gravitational force
between a sink particle, its host cell, and its neighboring cells, we
subdivide each cell into $8^3$ identical point particles, as described
in section~\ref{accretezone}. We set the force between the cell and
the sink particle equal to the sum of the forces between the sink
particle and the $8^3$ point particles.

The second step is to update the positions and momenta of the
particles including the effects of particle-particle
interactions. We handle this step separately because particles may
occasionally pass close to one another, within a few cell
spacings. When this happens, a time step chosen via equation
(\ref{partcourant}) may allow the particles to change their separation 
by a significant amount in a single update. In this case, a simple
first- or second-order position and momentum update will not provide
an accurate integration of the particle orbits. However, since
computing gas updates is far more expensive than computing particle 
updates, we do not wish to reduce our overall time step and compute
more gas updates. Instead, we integrate the particles forward through
a time $\Delta t$ using a Bulirsch-Stoer method with an adaptive
time step and error control \citep{numericalrecipes}. During this
integration, we consider only particle-particle gravitational
interactions, which, as with the particle-gas interactions, we compute
via a direct sum. Unlike with particle-gas interactions, we do not use 
a softened force law for particle-particle interactions. We have
tested this method by placing two particles in an extremely eccentric
orbit ($e$ = 0.998), where the closest approach of the particles is $\sim
\Delta x/100$, and the gas density and temperature are chosen so that
accretion is negligible. In this test we found that after several
orbits the semi-major axis was conserved to $\sim 1\%$ and the
eccentricity to $\sim 0.1\%$.

\section{Tests of the Methodology}
\label{tests}

\subsection{The Collapse of an Isothermal Sphere}

We have tested this method against the analytic solution for the
collapse of an isothermal sphere with a density 10\% above the
critical value \citep{shu77}. We consider a 1 $M_{\odot}$ sphere of
molecular H and He mixed in the standard cosmic abundance with a sound
speed of 0.18 km/s, appropriate for $\sim 10$ K gas. To avoid the
singular initial configuration of the Shu solution, we set our initial 
density and velocity profile to their analytic values after a time $t
= 1.3 \times 10^{12}$ s, when the expansion wave has propagated $2.4
\times 10^{16}$ cm, or 8 cells, from the origin. There are 64 cells in
the radius of the sphere, and the sphere is motionless and centered on 
the origin. To ensure that there are no problems when the sink
particle crosses cell boundaries, we ran two more tests with
identical setups except that we gave the isothermal sphere an initial
uniform velocity relative to the grid. In one test we used an
advection velocity of half the sound speed, and in the other twice the 
sound speed.

\begin{figure}
\plotone{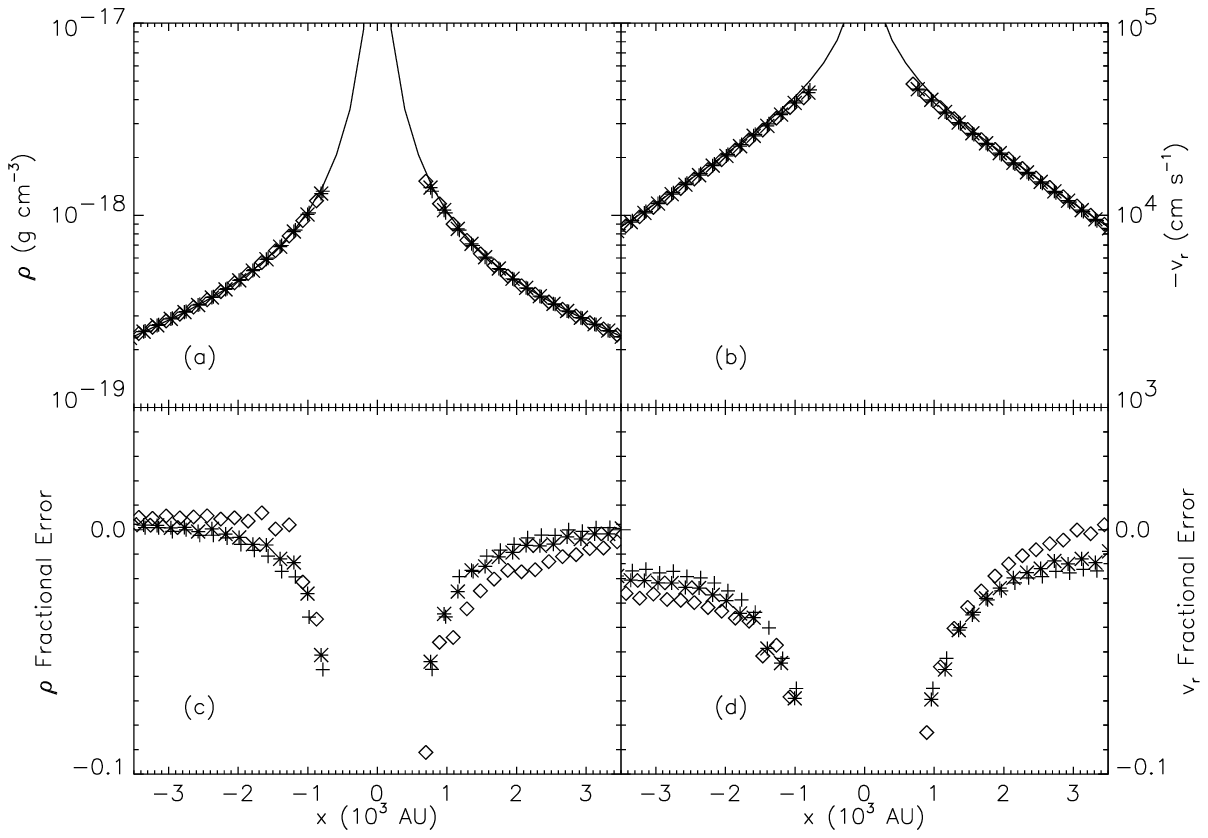}
\caption{\label{isosphere1}
The plots show the results of our isothermal sphere test. Panels (a)
and (b) show density and velocity versus radial distance, and panels
(c) and (d) show fractional error in density and velocity. In the
upper panels, the analytic solution is the solid line. In all panels
the crosses show values from the unadvected run, the asterisks show
values from the run advected at Mach 0.5, and the diamonds show values
from the run advected at Mach 2. The simulation data stop at the edge
of the sink region. All the plots are at time $t=2.0\times 10^{12}$ s.
}
\end{figure}

\begin{figure}
\plotone{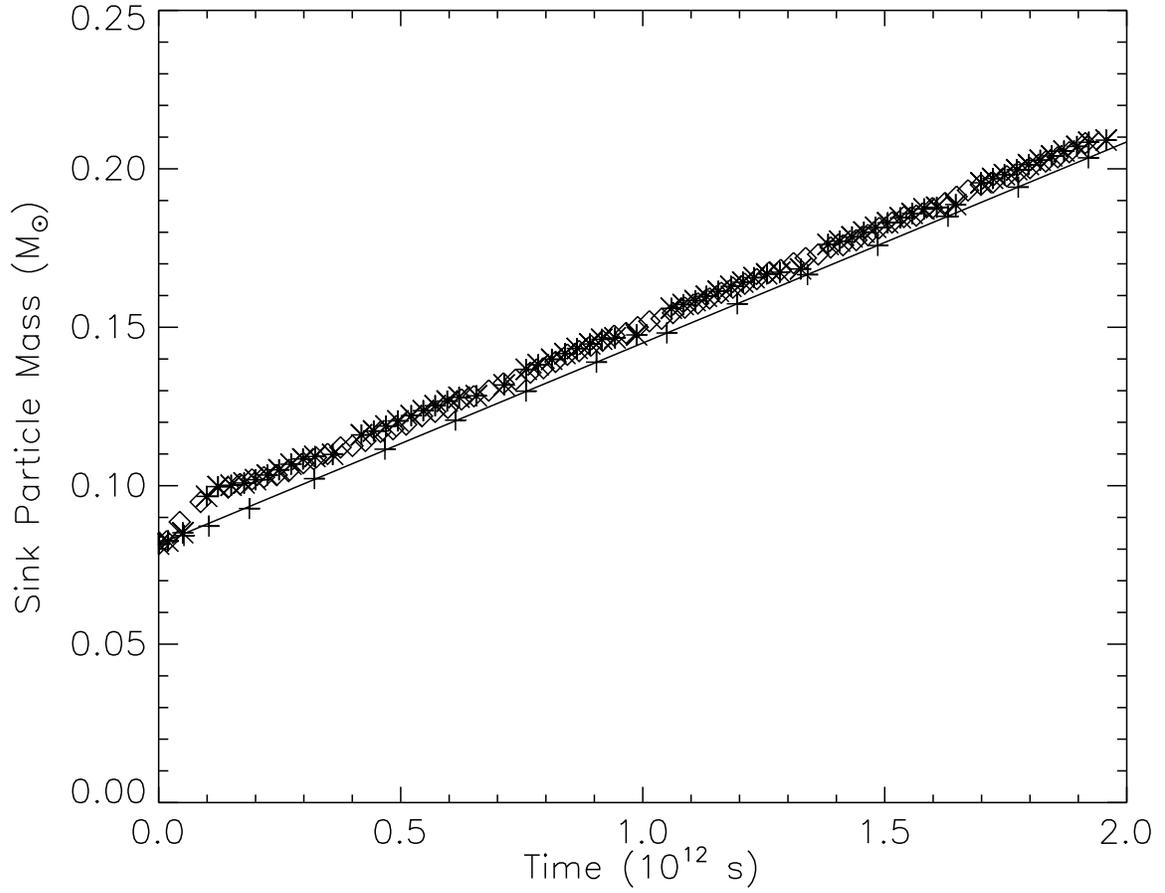}
\caption{\label{isosphere2}
The plot shows the mass of the sink particle versus time. The solid
line is the theoretical result. The crosses show values from the
unadvected run, the asterisks show values from the run advected at
Mach 0.5, and the diamonds show values from the run advected at Mach 2.
}
\end{figure}

The density and velocity profiles that we find from these tests are
shown in figure \ref{isosphere1}, and the mass of the sink particle
versus time is shown in figure \ref{isosphere2}. The calculation
reproduces the analytic solution extremely well. Errors in the
velocity and density profiles are a few percent in the cells adjacent
to the accretion region, dropping rapidly to $\sim 1\%$ as one moves further
away. The advected cases show a small initial transient during which the
accretion rate deviates from the theoretical value, but after a time
of less than the sound-crossing time of the accretion region the flow
settles into a steady state. Once this happens, the error in the
accretion rate is $\sim 1$\% in all three runs.

\subsection{Bondi Accretion}
\label{bondisim}

To test how well our sink particle formalism works
in cases where our resolution is marginal, so that $r_{\rm BH} \sim
\Delta x$, we ran a series of simulations of simple Bondi
accretion using a variety of values for $r_{\rm BH}/\Delta x$. In each 
case we use a sphere of mixed H and He at 10 K with a radius of
$1.21\times 10^{19}$ cm. For 32 cells across the radius of the sphere, 
and a 1 $M_{\odot}$ central star this gives $\Delta x = r_{\rm
BH}$. We initialize the density and velocity profile of the sphere to 
the analytic solution for the Bondi problem and allow the
calculation to run until the accretion rate reaches steady state. We
then repeat the calculation with different sink particle masses and
compare the accretions rates to the theoretical predictions. The
results are shown in Table \ref{bonditab}.

As the table indicates, there can be a substantial error in the
accretion rate when the Bondi radius of the sink particle is
comparable to a cell spacing, but the error drops off rapidly as the
Bondi radius either increases or decreases. The peak error seems to
occur when the Bondi radius is approximately equal to the radius of
the accretion zone. For a factor of 10
difference in $r_{\rm B}$ and $r_{\rm acc}$ the error is $\sim 1\%$,
while for even a factor of $10^{1/2}$ difference the error is only
$\sim 10\%$ or less. We experimented with several factors of order unity in
equation (\ref{rhobar}), where we set $\rho_{\infty}$, to see if we
could decrease the error. We found that 1.2 gave the best results, but 
that other factors between 0.5 and 2.0 increased the error in the
accretion rate by no more than $\sim 10\%$.

This substantial error when $r_{\rm B} \approx r_{\rm acc}$ is not
particularly surprising. The Bondi radius is the point at which the
flow transitions from subsonic to supersonic. If that is equal to the
accretion radius, inside which we are artificially altering the
physics, then the transition is not well resolved, and the density and
velocity in cells near the sink particle will have substantial
errors. These errors lead the code to set an incorrect accretion rate,
which in turn compounds the problem by setting a back-pressure that is
too large. When the flow near the sink particle is either subsonic or
supersonic, this problem does not occur and the errors are far
smaller. Nonetheless, this provides an important caveat to our method:
one ought not use it in cases where the flow at the sink particle is
transitioning from subsonic to supersonic.

\subsection{Bondi-Hoyle Accretion}
\label{bondihoyle}

We tested the ability of our sink particle method to handle accretion
from a moving medium by simulating Bondi-Hoyle accretion. We place a
1 $M_{\odot}$ sink particle in an initially uniform gas, composed of a
standard interstellar mix of H and He (mean particle mass of
$2.33m_p$) at 10 K, with a 
density of $10^{-25}$ $\mbox{g}\;\mbox{cm}^{-3}$. The gas flows past
the sink particle at Mach 3. We use inflow boundary conditions in the
upstream direction and outflow boundaries downstream. 
The resolution of the finest cells in the calculation is $7.4\times
10^{14}\;\mbox{cm} = \rbh/50$, where $\rbh$ is as defined by
equation (\ref{rbh}). 

Using equation (\ref{accrate}), the
expected accretion rate is $\dot{M}_{\rm BH} = 1.7\times10^{-12}$
$M_{\odot}\;\mbox{yr}^{-1}$. However, this simulation is in a regime
where the interpolation formula works poorly. \citet{ruffert96}
performed a
simulation very similar to ours (the run labeled GS, which has a
small accretor, gas with polytropic index $\gamma=1.01$, $\calm=3$),
and found an accretion rate of $2.0\times10^{-12}$
$M_{\odot}\;\mbox{yr}^{-1}=1.17\,\dot{M}_{\rm BH}$, where we have scaled
their dimensionless
result to our dimensional parameters. \citet{ruffert96} also found
that both the accretion rate and the flow pattern are
time-dependent, and that the flow pattern shows substantial deviations
from axial symmetry on scales comparable to $r_{\rm BH}$. The
mechanism for disrupting steady, symmetric flow is not fully
understood, but \citet{foglizzo99} suggest Rayleigh-Taylor and
Kelvin-Helmholtz instabilities at the shock front as the probable
cause.

\begin{figure}
\plotone{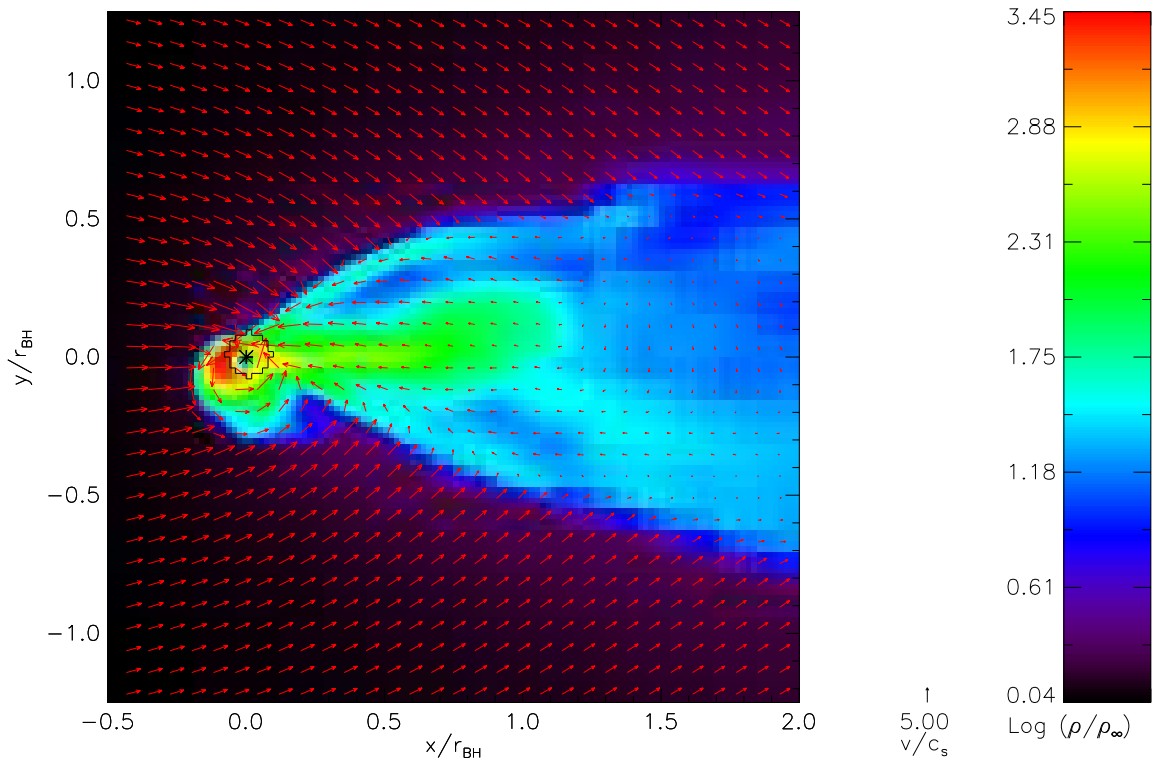}
\caption{\label{bondihoyle.fig}
The plot shows the density and velocity field in the XY plane at
$t\approx 12 \tbh$. The sink particle is at the origin, and the white 
border indicates the boundary of the accretion zone. One can clearly see
the gravitational focusing of streamlines into a shock behind the accretor 
that is characteristic of Bondi-Hoyle accretion.
}
\end{figure}

\begin{figure}
\plotone{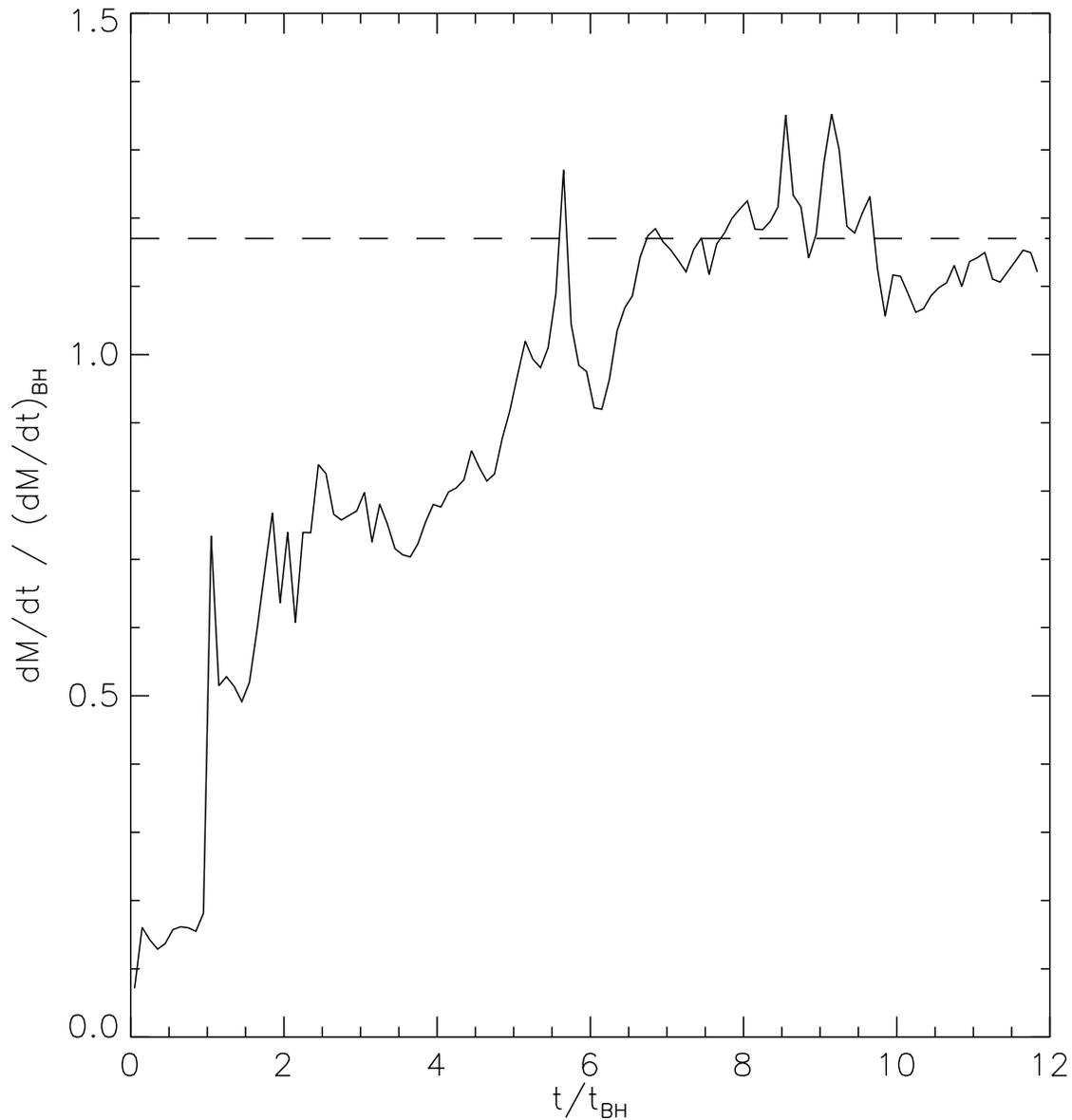}
\caption{\label{bondihoyle.accrate}
The plot shows the accretion rate as a function of time in our
simulation of Bondi-Hoyle accretion. The points are sampled at
intervals of $\tbh/10$. Time is plotted in units of $\tbh=6.35\times
10^4$ yr, and accretion rate is plotted in units of $\dot{M}_{\rm
BH}=1.7 \times 10^{-12}\;M_{\odot}\mbox{ yr}^{-1}$. The horizontal
dashed line is the accretion rate predicted by \citet{ruffert96}.
}
\end{figure}

Figure \ref{bondihoyle.fig} shows the system near the end of our
simulation, and Figure \ref{bondihoyle.accrate} shows the accretion
rate as a function of time. As expected, the flow is time dependent
and unstable, with no axial symmetry. The accretion rate requires
several Bondi-Hoyle times, defined by $\tbh\equiv\rbh/c_s=6.35\times
10^4$ yr, to reach an
equilibrium value, and even then it shows substantial
fluctuations. The accretion rate appears to reach equilibrium after
$\sim 6\tbh$; the average accretion rate after that point is
$2.0\times 10^{-12}\; M_{\odot}\;\mbox{yr}^{-1}=1.17\,\dot{M}_{\rm BH}$, in
agreement with the results of \citet{ruffert96}.

This test demonstrates that our method produces correct results for
non-symmetric, time-dependent flows. It also illustrates an important
point regarding our accretion formula, equation (\ref{accrate}): the
method is self-correcting, and thus the exact details of the accretion
formula do not make much difference as long as the Bondi-Hoyle radius
is well-resolved. The formula uses our best guesses for $v_{\infty}$
and $\rho_{\infty}$ based on the characteristics of the flow; however,
it would be unreasonable to expect our method to correctly guess these
values to the level of accuracy necessary for the formula to reproduce
the \citet{ruffert96} result. Instead, the accretion rate is
ultimately dictated by the rate at which the ordinary, unaltered
hydrodynamics of our simulation brings gas into the accretion
region. If equation (\ref{accrate}) sets an accretion rate that is too
low, gas will enter the region faster than it is removed by accretion,
so our guess for $\rho_{\infty}$ will increase and the sink particle
will consume gas more quickly. The opposite effect happens if equation
(\ref{accrate}) dictates that gas be removed too quickly. Thus, our
accretion rate is self-correcting.

\subsection{Rotating Flows}
\label{rotflow}

\subsubsection{Sink Particle Results}

Finally, to ensure that our method does not cause artificial accretion 
or angular momentum transport, we tested the evolution of a disk
around a sink particle. We place a 1 $M_{\odot}$ sink particle at the
center of a thin 10 K gas disk with a radius $r_0=2\times
10^{15}\;\mbox{cm}$ and a power-law surface density
profile $\Sigma = \Sigma_0 (r_0/r)^{k_{\rho}}$, with
$\Sigma_0=0.1\;\mbox{g}\;\mbox{cm}^{-2}$,
and $k_{\rho}=1$. The disk is in Keplerian 
rotation, and the grid is chosen so the finest cells are $2.1\times
10^{13}\;\mbox{cm}$ in length. We simulated the evolution of the
disk for more than 100 orbital periods at the edge of the accretion
region.

In analyzing this test, it is crucial to separate the effects of the
sink particle from those of ordinary numerical viscosity. As one
approaches the sink particle, the circles in which the gas is flowing
are resolved by fewer and fewer cells. Numerical viscosity therefore
becomes significant, and will cause angular momentum transport. This
effect causes evacuation of the gas in the disk near the sink
particle, with some of the gas falling towards the center and the rest 
pushed outwards. \citet{nelson2000} and \citet{okamoto2003} report
an analogous phenomenon in SPH calculations, as do
\citet{kuznetsov1999} in a 3D Eulerian code gridded inhomogenously in
spherical coordinates. To disentangle this
effect from possible artificial angular momentum transport due to the
sink particle, we ran two identical simulations. In one, we used the
standard sink particle. In the other, we disabled accretion onto the
sink particle. In this latter case, because the sink particle is far
more massive than the disk and thus does not move significantly during 
the calculation, the sink particle's sole effect is to impose a point
gravitational potential.

\begin{figure}
\plotone{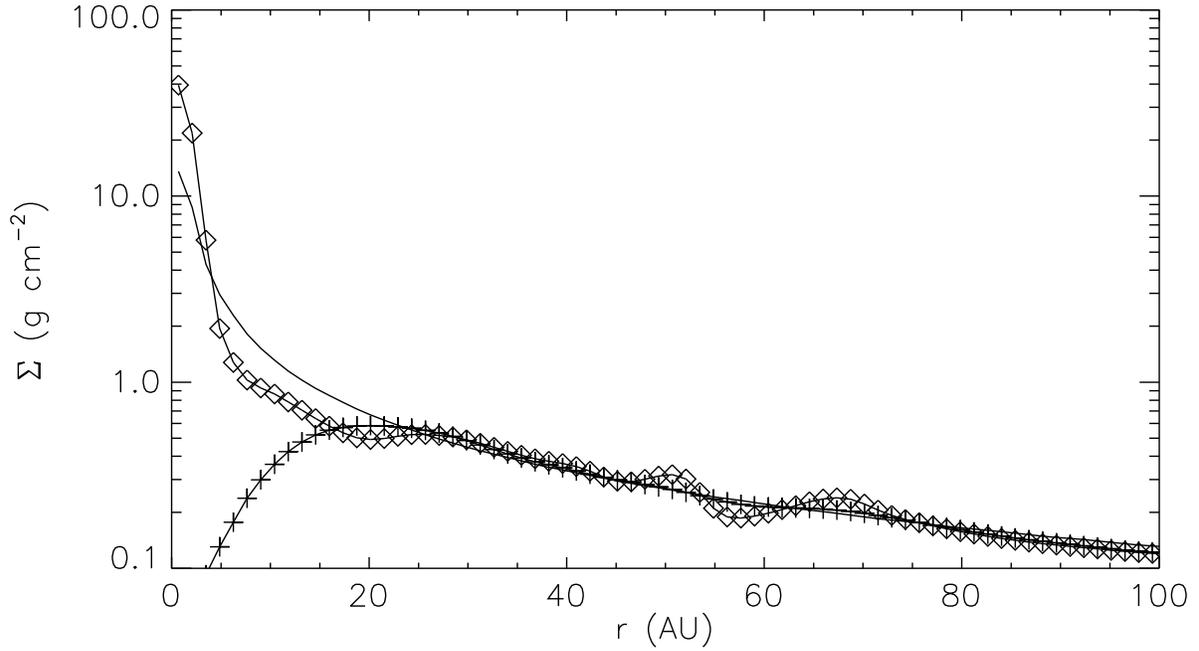}
\caption{\label{colprof}
The plot shows the azimuthally-averaged surface density versus radius
after 50 orbital periods at the edge of the accretion region,
$r=5.55\mbox{ AU}$, in simulations of disk evolution around a sink
particle. The line
marked with diamonds shows the run without sink particle accretion,
the line marked with crosses shows the run with accretion, and the
unmarked line shows the initial surface density profile. The interval
in radius between data points is one cell. 
}
\end{figure}

Figure \ref{colprof} shows the surface density versus radius at nearly
identical times in the two calculations. In the simulation without
accretion, there is a clear dip in the surface density between about 5
and 20 AU. Some of this gas has fallen into an unresolved hydrostatic
object extending a few cells from the origin, leading to a density
enhancement there. The rest has been pushed out further into the disk, 
leading to the alternating enhanced and diminished density between 45
and 70 AU. Examination of surface density profiles at other times shows
that this phenomenon is a wave moving out from the origin, a result
of material being pushed away from the sink particle by numerical
viscosity.

In the calculation with sink particle accretion, the column 
density matches the initial surface density well except within about 20 AU
of the sink particle, where it falls sharply. In this case, the gas
being evacuated from the low resolution region around the sink
particle has mostly been accreted. There is a slight density
enhancement just outside the evacuated region, but it is smaller both
in magnitude and in extent than in the calculation with no
accretion. Similarly, the wave caused by material pushed outwards from
the origin is far smaller in this simulation In both calculations the
evacuated region is approximately the same size. The small decrease in
density at radii $> 80$ AU apparent in both simulations is a
result of the pressure boundary conditions on the disk, and is
unrelated to the sink particle.

\begin{figure}
\plotone{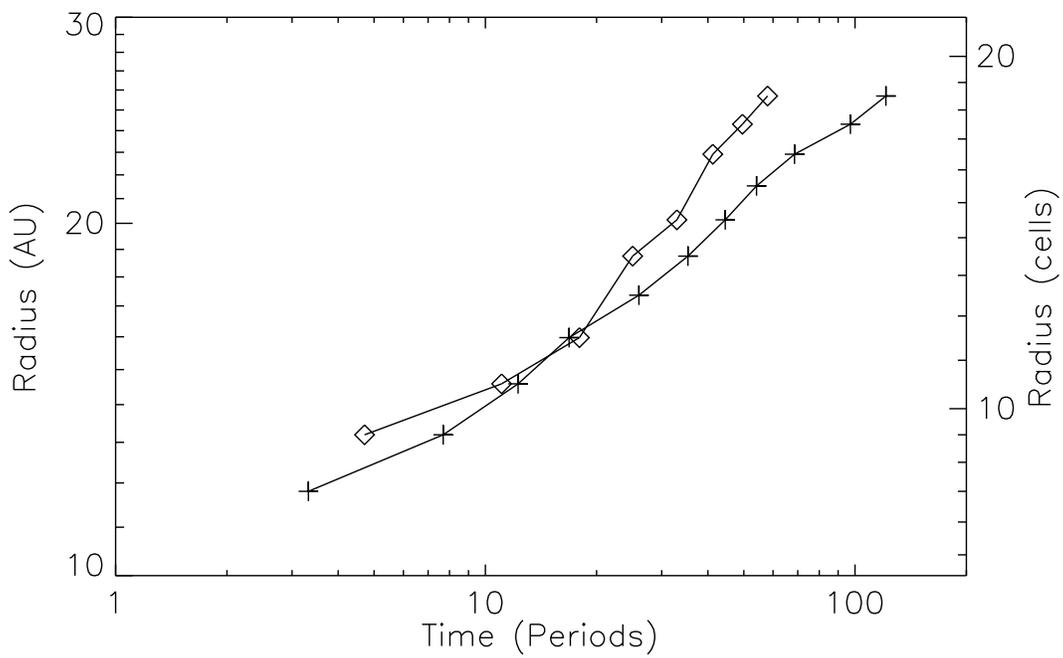}
\caption{\label{evacrad}
The plot shows the radius of the evacuated region versus time in
simulations of disk evolution around a sink particle. The line
marked with diamonds shows the run without sink particle accretion,
and the line marked with crosses shows the run with accretion.
}
\end{figure}

Figure \ref{evacrad} shows the size of the evacuated region of the
disk versus time. We define the edge of the evacuated region as the
smallest radius for which the surface density is 90\% or more of the
initial surface density at that radius. In the simulation without
accretion, we exclude the inner 4 cells where the hydrostatic gas has
accumulated. As the plot shows, the radius of the evacuated region is
slightly less in the run with sink accretion than without. This
indicates that the sink particle is not causing enhanced accretion or
angular momentum transport. The 
sink particle may actually lead to better results by preventing
material that is artificially pushed outward by numerical viscosity
from escaping the accretion region and affecting the rest of the
calculation.

\subsubsection{Estimating $\alpha$ Due to Numerical Viscosity}
\label{diskalpha}

As a side note, in the calculation with
accretion, a power-law fit to the radius of the evacuated region
versus time gives
\begin{equation}
\label{evac}
\frac{r_{\rm evac}}{\Delta x} = 6.1 \; \left[\frac{
\Omega\left(r_{\rm acc}\right) t}{2\pi}\right]^{0.23},
\end{equation}
where $\Omega\left(r_{\rm acc}\right)$ is the angular velocity of the
disk at the edge of the accretion region; recall that we use $r_{\rm
acc}=4\Delta x$ in this work. We can interpret this
relation as a resolution requirement for the number of cells $r_{\rm
evac}/\Delta x$ as a
function of total run time $t$ that we need to ensure that numerical
viscosity does not affect structures at a specified
distance $r_{\rm evac}$ from the sink particle. We can also compute an
approximate
viscosity parameter $\alpha$ for this effect. For an isothermal
Keplerian disk orbiting around an object of mass $M$, the accretion
time scale at a distance $r$ from the central object is
\begin{equation}
t_{\rm acc} \approx \frac{r^2}{\nu},
\end{equation}
where $\nu$ is the kinematic viscosity \citep{lyndenbell74}. The
standard $\alpha$ parameter is defined \citep{shakura1973} so that
\begin{equation}
\nu = \frac{\alpha c_s^2}{\left| r \frac{d\Omega}{dr}\right|} =
\frac{\alpha c_s^2}{\frac{3}{2}\sqrt{\frac{GM}{r^3}}},
\end{equation}
where $c_s$ is the sound speed in the disk and $\Omega$ is the disk
angular velocity. Eliminating $\nu$,
\begin{equation}
\label{alpha}
\alpha \approx \frac{3}{2} \frac{\left(G M r\right)^{1/2}}{t_{\rm acc}
c_s^2} = \frac{3}{2} \frac{GM}{r c_s^2} \frac{1}{\Omega t_{\rm acc}}.
\end{equation}
We can identify the evacuation radius and time of equation
(\ref{evac}) with the accretion radius and timescale of equation
(\ref{alpha}). Doing so, we find that the numerical viscosity as a
function of radius is
\begin{equation}
\label{alphaeq}
\alpha \approx 78 \; \frac{GM}{r c_s^2} \; \left(\frac{r}{\Delta
x}\right)^{-2.85} = 78\; \frac{r_{\rm B}}{\Delta x} \;
\left(\frac{r}{\Delta x}\right)^{-3.85},
\end{equation}
where $r_{\rm B}\equiv GM/c_s^2$ is the standard Bondi radius. Note
that the exponent of the relation for $\alpha$ is $-2.85$ rather than
$-4.35=-1/0.23$ because $\Omega\propto r^{-3/2}$. To give some feel
for the consequences of this relation, for $r_{\rm B}/\Delta x=10$ the
numerical viscosity drops to $\alpha=0.01$ at $r=18.6\;\Delta
x$.

Equation (\ref{alphaeq}) provides a useful resolution
requirement for hydrodynamics codes involving disks, as one can only
believe the results of a disk simulation in those regions with
resolution high enough that the effective $\alpha$ due to numerical
viscosity is much smaller than the $\alpha$ that arises from whatever
sources of physical viscosity are present. The scaling of $\alpha$
with $r$, $\rb$, and $\Delta x$
depends only on geometry and on the physics of $\alpha$-disks, and so
is likely to be about the same in any code using Cartesian geometry;
the constant of proportionality, $78$ for our code, likely depends on
the hydrodynamic algorithm. Hydrodynamic codes using different grid
geometries, on the other hand, would likely have very different
exponent in (\ref{alphaeq}). The rather large value of $\alpha$ we
find is for a Cartesian grid. A disk centered on the origin of a
cylindrical or spherical grid would probably show much less numerical
viscosity.

\section{From Bondi to Bondi-Hoyle Accretion}
\label{bondi2bh}

\subsection{Background}

While there have been extensive studies of Bondi-Hoyle accretion for
flow with uniform velocities or smooth velocity gradients
\citep{ruffert94a, ruffert94b, ruffert94c, ruffert95a, ruffert95b,
ruffert96, foglizzo97, foglizzo99, foglizzo2002}, in many cases
accretion occurs in a supersonically turbulent medium in which there
are numerous shocks present. An example of accretion in a shock-filled 
medium is the star formation process. Observations of star forming
regions show
they they are turbulent, with complex morphologies and
velocity structures. Non-thermal linewidths within star-forming
regions range from transonic for the length scale of a core that forms
a single star or small multiple system to highly supersonic on the
scale of entire molecular
clouds, with a power-law relation between linewidth and size in
between \citep{larson81, ossenkopf2002}. Simulations of turbulence
in star forming cores show that turbulent motions are able to
reproduce the molecular line emissions, aspect ratios, linewidth
gradients, and linewidth-size relations \citep{klein2003, padoan2003}.

It is therefore interesting to consider what happens when a shock
rolls over an accreting object, such as a protostar in a molecular
cloud clump. Initially,
the accretion rate and density and velocity profiles will look like
standard Bondi accretion; after long times, they will be appropriate
for Bondi-Hoyle accretion. We are interested in studying the details
of the transition and the time-dependence of the accretion rate.

\subsection{The Simulation}

\begin{figure}
\plotone{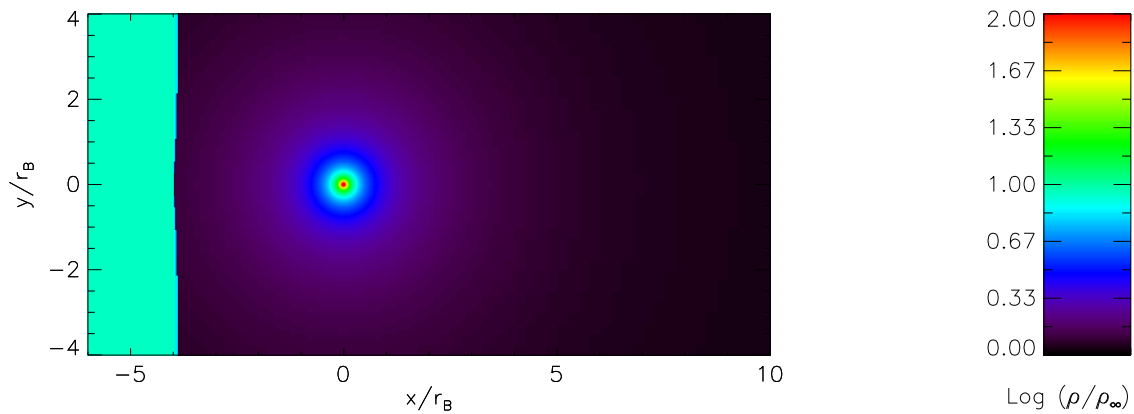}
\caption{\label{bondi2bh.init1}
The plot shows the logarithm of density for the initial
configuration of the Bondi to Bondi-Hoyle accretion problem. The image 
is a slice through the equatorial plane. We do not show the accretion
region or sink particle because they are too small to see clearly.
The plotted density range
has been truncated at the top to bring out lower density
features. The small irregularity visible in the shock front is a
result of varying resolution: the shock is wider further from the $x$
axis because it is spread over $\sim 3$ cells and the cells are larger 
further from the axis. The Bondi radius is $\rb = 0.12$ pc,
$\rho_{\infty}=10^{-25}\mbox{ g cm}^{-3}$.
}
\end{figure}

\begin{figure}
\plotone{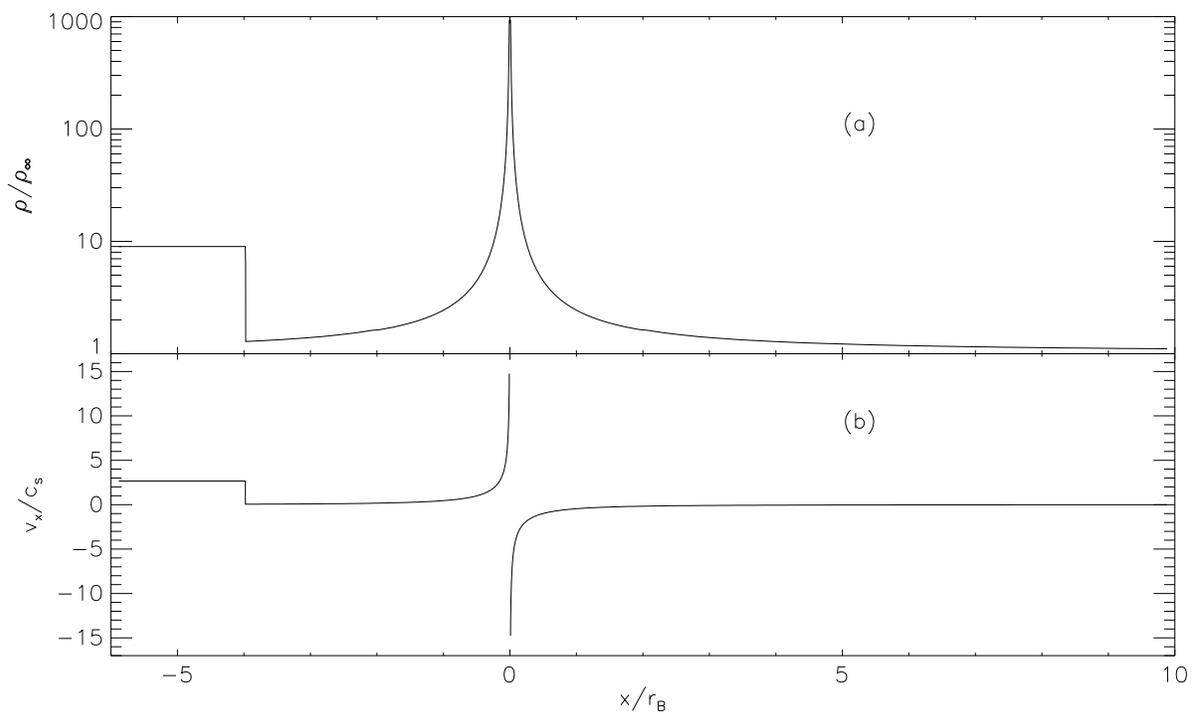}
\caption{\label{bondi2bh.init2}
Panel (a) shows the density panel (b) shows
the $x$ velocity along the $x$ axis 
in the initial configuration of our Bondi to
Bondi-Hoyle accretion problem. The
Bondi radius, sound speed, and background density are $\rb = 0.12$ pc,
$c_s=0.19$ km/s, and $\rho_{\infty}=10^{-25}\mbox{ g cm}^{-3}$.
}
\end{figure}

We simulate a Mach 3 shock impacting an accreting particle. We place a
sink particle with a mass of $M_{\odot}$ in a gas of H and He with a
temperature of 10 K. The Bondi radius is therefore $\rb =
0.12$ pc. Figures \ref{bondi2bh.init1} and \ref{bondi2bh.init2} show
the initial 
configuration of the problem. We divide the computational domain into
two regions. For $x>-4 \rb$, the gas initially has a density and
velocity profile given by the analytic solution for Bondi accretion
with $\rho_{\infty}=10^{-25}\mbox{ g}\mbox{ cm}^{-3}$. This density is
unrealistically low for a real star-forming region; we choose it
because we wish to ensure that the particle does not accrete enough
to substantially change its mass over the course of the
simulation. We are neglecting the self-gravity of the gas, and thus,
as long as the sink particle's mass changes negligibly, the density is
not a relevant parameter of the problem and may be scaled to an
arbitrary value. For $x<-4 \rb$, we generate post-shock conditions
appropriate for an isothermal shock of Mach number $\calm=3$ moving in 
the $+x$ direction into a region with density $\rho_{\infty}$ that is
at rest. Thus, for $x<-4\rb$, we set  $\rho=\calm^2\rho_{\infty}$,
$\mathbf{v}=\left(\calm - 1/\calm\right) c_s \mathbf{\hat{x}}$, where
$c_s$ is the sound speed. 

The domain of the computation goes from
$-4\rb$ to $4\rb$ in the $y$ and $z$ directions, and from $-6 \rb$ to
$10 \rb$ in the $x$ direction. The boundary conditions are
inflow/outflow in every direction. We choose refinement criteria to
guarantee that the region around the sink particle is always resolved
such that, at a distance $r$ from the sink particle, $r/\Delta x \ge
32$. This continues to a maximum resolution of $\Delta x = \rb/512 =
50$ AU. Note that, for $\calm=3$, $\rb = 10r_{\rm BH}$, so the maximum
resolution is 51 cells per Bondi-Hoyle radius.

Before the shock reaches the sink particle, the flow is pure
Bondi accretion; for the parameters we use in the simulation, the
accretion rate should be $\dot{M}_{\rm B} = 5.9\times 10^{-11}$ $M_{\odot}
\;\mbox{yr}^{-1}$. Long after the shock passes the particle, the flow
should be Bondi-Hoyle accretion with $\calm=3$ and a background
density of $\calm^2 \rho_{\infty} = 9\times 10^{-25} \mbox{
g}\mbox{ cm}^{-3}$. Based on our results from Section~\ref{bondihoyle} 
and those of \citet{ruffert96}, the mean accretion rate should then be 
$\dot{M}_{\rm BH} = 1.8\times 10^{-11}$ $M_{\odot}\;\mbox{yr}^{-1}$.

\subsection{Results}

\begin{figure}
\plotone{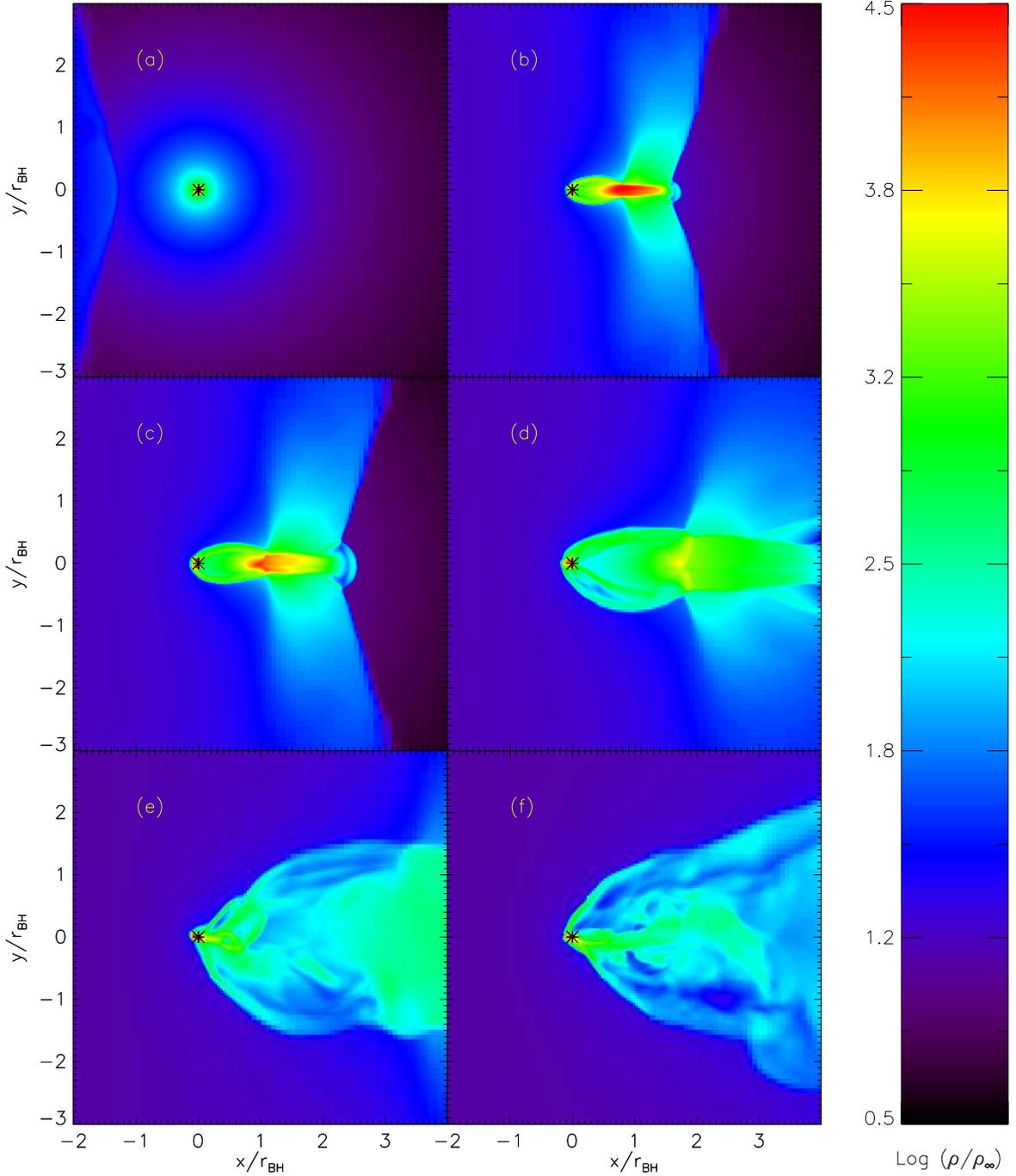}
\caption{\label{bondi2bh.time}
The panels show a series of slices through the equatorial plane. The
density range has been truncated to bring out detail. The asterisks
indicate the position of the sink particle; we
do not show the accretion region because it is too small to see
clearly. We have truncated the density range at the top and the bottom 
to bring out details.
Note that, in contrast with Figures \ref{bondi2bh.init1} and
\ref{bondi2bh.init2}, the
length unit of these plots is $\rbh=0.012\mbox{ pc}=\rb / 10$. The
times shown in each panel are: (a) $-0.25 \;\tbh$; (b) $1.6 \;\tbh$; (c)
$1.9 \;\tbh$; (d) $2.7 \;\tbh$; (e) $4.3 \;\tbh$; (f) $5.9 \;\tbh$, where
$\tbh = 6.35\times 10^4$ yr.
}
\end{figure}

Figure \ref{bondi2bh.time} shows a series of snapshots of the
simulation. As one might expect, the configuration does not change
much until the shock approaches within a distance $\sim \rb$ of the
particle. At that point, the shock starts to bow, with the part along
the axis the furthest forward as it propagates into infalling
gas. Panel (a) shows this effect. Panel (b) shows that as the
shock passes the particle, a dense cylinder of shocked material
accumulates where flows of gas swept up in the shock converge and
shock again. This is the beginning of the converging streamlines and
Mach cone that are characteristic of Bondi-Hoyle accretion. In panel
(c), the dense cylinder is showing the first signs of the destruction
of axial symmetry. In panel (d), the asymmetry is increasing. By
panel (e), within one Bondi-Hoyle radius of the sink particle there is
a well-developed Mach cone, which is becoming turbulent in its
interior. In panel (f), the Mach cone extends the full length of the
range we plot, and appears to be fully turbulent in its interior.
However, in panel (f) the density
of material in the Mach cone is still inflated due to the presence of
material that was part of the Bondi flow and has been swept up by the
shock. The densities in the cone undergo a slow decline until reaching
a steady state; however, the morphology of the cone does not change
further.

\begin{figure}
\plotone{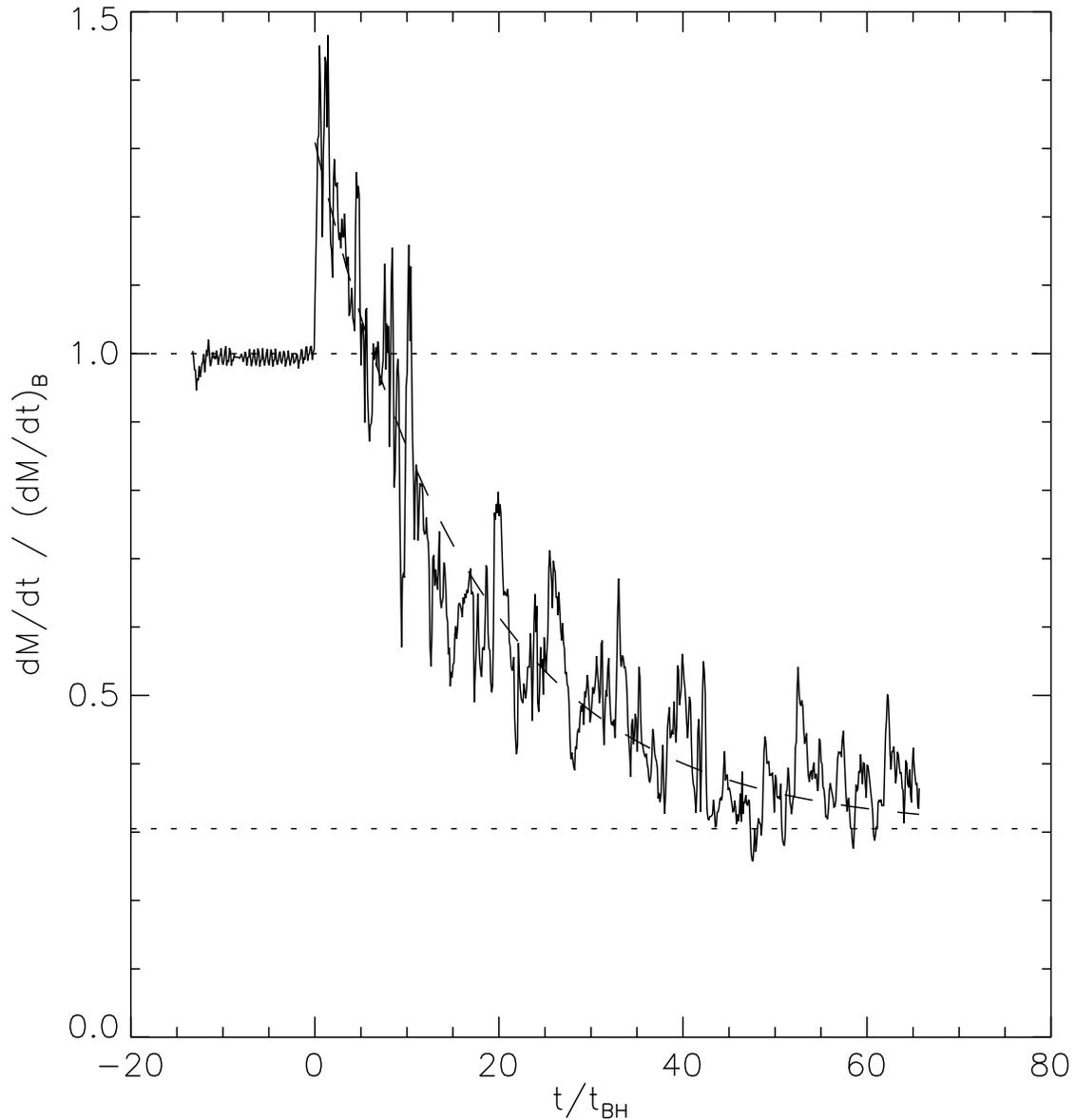}
\caption{\label{bondi2bh.accrate}
The solid curve shows the accretion rate onto the sink particle versus time,
sampled at intervals of $\tbh/10$. The short-dashed horizontal lines show
the predicted Bondi (upper line) and Bondi-Hoyle (lower line) accretion
rates. The long-dashed line shows our exponential fit to the accretion
rate after the shock hits the sink particle. Time is plotted in units
of $\tbh=6.35\times 10^4$ yr, and accretion rate is plotted in units
of $\dot{M}_{\rm B} = 5.9\times 10^{-11}\;M_{\odot}\mbox{ yr}^{-1}$.
}
\end{figure}

Figure \ref{bondi2bh.accrate} shows the accretion rate as a function
of time. As the plot shows, the accretion rate is initially flat and
in good agreement with the predicted value for Bondi accretion. When
the shock hits the sink particle at $t=0$, the accretion rate
immediately increases by $\sim 50\%$ as the particle starts accreting
high density shocked material. The accretion rate then begins to fall
off until it approaches its equilibrium value. After some
experimentation, we found that the overall shape curve is reasonably
well-fit by an exponential of the form
\begin{equation}
\dot{M} = \dot{M}_{\rm BH} + \left(\dot{M}_0 -\dot{M}_{\rm BH}\right)
\exp\left(-\frac{t}{t_{\rm trans}}\right),
\end{equation}
where $\dot{M}_0$ is the accretion rate immediately after the shock
hits the sink particle and $t_{\rm trans}$ is the characteristic
timescale for the accretion rate to transition from Bondi to
Bondi-Hoyle. Our best-fit values for this case are $\dot{M}_0 =
7.7\times 10^{-11}\;M_{\odot}\;\mbox{yr}^{-1} = 1.3 \; \dot{M}_{\rm
B} = 4.3 \; \dot{M}_{\rm BH}$ and $t_{\rm trans} = 1.1\times
10^7\mbox{ yr} = 1.7 \; \tb =
17 \; \tbh$, where $\tb=\rb/c_s=6.35\times 10^{3}$ yr and $\tbh=
\rbh/c_s=6.35\times 10^{4}$ yr.

The characteristic timescale for the transition is closer to 
the Bondi time than the Bondi-Hoyle time. This is
not surprising in retrospect. Before the shock hits,
gas out to a distance $\sim \rb$ from the sink
particle is inflowing supersonically. In a Bondi-Hoyle flow, gas at
distances $\sim \rb \gg \rbh$ does not develop supersonic inflow
velocities because it does not spend enough time near the sink
particle. However, even after it is shocked, the gas that was
originally part of the Bondi flow retains its supersonic infall speed
and is therefore likely to find its way down to
the sink particle. Since the gas is
coming from a distance of order $\rb$, its characteristic timescale
to reach the sink particle is of order $\tb$. Once all the leftover gas
that was part of the Bondi flow out to $\sim \rb$ has drained onto the 
accreting particle, the accretion rate drops down to what one would
expect for pure Bondi-Hoyle flow. As a result of this effect, the
particle accretes an additional amount of mass
\begin{equation}
\Delta M \approx \int \left(\dot{M}-\dot{M}_{\rm BH}\right) dt
\approx \left(\dot{M}_0-\dot{M}_{\rm BH}\right) t_{\rm trans} 
\approx 56 \; \dot{M}_{\rm BH} \tbh
\end{equation}
beyond what it would have accreted if the accretion rate had instantly 
shifted from Bondi to Bondi-Hoyle.

\section{Summary}
\label{conclusion}

We have demonstrated a new method for including moving, Lagrangian
sink particles in an Eulerian hydrodynamics code. Our method shows
excellent agreement with analytic results for a number of test
problems, even in  regimes where our formula for the accretion rate is
at best a guess. In the process of testing the behavior of our sink
particle in the presence of rotating flows, we develop a method 
to parameterize the effects of numerical viscosity in disk
simulations, and use it to compute $\alpha$ for our Cartesian AMR code.
This method provides a
resolution requirement for future disk simulations.

We have also solved a simple example problem using our sink particle
method, and shown that it produces useful results in that case. 
We have discovered one limit of our method: that sink particles
can produce significant errors in the 
accretion rate when the flow is transitioning from subsonic to
supersonic at the accretion radius. It is not clear if SPH sink
particles also encounter difficulty in this regime; we have not found
any tests in the literature that address the point. Regardless, with
AMR one can easily avoid the regime where $\rb\sim r_{\rm acc}$, since
once may simply increase the resolution temporarily until the sink
particle accretes enough mass that $\rb> r_{\rm acc}$.

The new technique will extend the range of Eulerian simulations,
allowing them to run for longer times on problems gravitational
collapse. Thus, it extends to Eulerian codes one of the heretofore
unique advantages of SPH, while retaining the accurate treatment
of shocks and radiative transfer capability which are found in
Eulerian approaches.

\acknowledgements The authors thank Fumitaka Nakamura and
Robert Fisher for useful discussions. This work was supported by:
NASA GSRP grant NGT 2-52278 (MRK); NSF grant
AST-0098365 (CFM); NASA ATP grant NAG 5-12042 (CFM and RIK);
the US
Department of Energy at the Lawrence Livermore National Laboratory
under contract W-7405-Eng-48. This research used resources of
the National Energy Research Scientific Computing Center, which is
supported by the Office of Science of the U.S. Department of Energy
under Contract No. DE-AC03-76SF00098, and the NSF San Diego
Supercomputer Center through NPACI program grant UCB267.

\clearpage

\clearpage 

\clearpage

\begin{deluxetable}{lrrrr}
\tablecaption{Simulated versus Theoretical Accretion Rates for Bondi
Accretion.\label{bonditab}}
\tablewidth{0pt}
\tablehead{
\colhead{$M_{\rm sink}$ ($M_{\odot}$)} & \colhead{$r_{\rm B}/\Delta_x$} &
\colhead{$\dot{M}_{\rm theor}$ ($M_{\odot}$/yr)} &
\colhead{$\dot{M}_{\rm sim}$ ($M_{\odot}$/yr)} &
\colhead{$\dot{M}$ error}}
\startdata
0.1 & 0.1 & $5.94\times 10^{-13}$ & $6.01\times 10^{-13}$ & $0.011$
\\
0.316 & 0.316 & $5.94\times 10^{-12}$ & $5.90\times 10^{-12}$ & $-0.006$
\\
1.0 & 1.0 & $5.94\times 10^{-11}$ & $5.21\times 10^{-11}$ & $-0.122$
\\
3.16 & 3.16 & $5.94\times 10^{-10}$ & $4.49\times 10^{-10}$ & $-0.244$
\\
10.0 & 10.0 & $5.94\times 10^{-9}$ & $5.76\times 10^{-9}$ & $-0.023$
\\
\enddata
\tablecomments{Column 1 shows the sink particle mass, column 2 shows the
ratio of the Bondi radius to a grid spacing, columns 3-4 show the
theoretical and simulated accretion rates, and column 5 shows the
fractional error in the simulated accretion rate.}
\end{deluxetable}

\clearpage 

\clearpage

\clearpage

\clearpage

\clearpage

\clearpage

\clearpage

\clearpage

\end{document}